\documentclass[conference]{IEEEtran}
\ifCLASSINFOpdf
\else
\fi

\usepackage{soul}
\usepackage{xcolor}
\sethlcolor{yellow}
\usepackage{graphicx}
\usepackage{hyperref}

\usepackage{url}

\begin{document}
%
\title{\textit{``I'm 73, you can't expect me to have multiple passwords''}: Password Management Concerns and Solutions of Irish Older Adults}

\author{\IEEEauthorblockN{Ashley Sheil\IEEEauthorrefmark{1},
Jacob Camilleri, Michelle O Keeffe,
Melanie Gruben,
Moya Cronin and
Hazel Murray\IEEEauthorrefmark{2}}
\IEEEauthorblockA{Munster Technological University, Co Cork, Ireland\\
Email: \IEEEauthorrefmark{1}ashley.sheil@mtu.ie,
\IEEEauthorrefmark{2}hazel.murray@mtu.ie
}}


\IEEEoverridecommandlockouts
\makeatletter\def\@IEEEpubidpullup{6.5\baselineskip}\makeatother
\IEEEpubid{\parbox{\columnwidth}{
		Symposium on Usable Security and Privacy (USEC) 2025 \\
		24 February 2025, San Diego, CA, USA \\
		ISBN 979-8-9919276-5-9 \\
		https://dx.doi.org/10.14722/usec.2025.23017 \\
		www.ndss-symposium.org, https://www.usablesecurity.net/USEC/
}
\hspace{\columnsep}\makebox[\columnwidth]{}}

\maketitle



\begin{abstract}
Based on Irish older adult's perceptions, practices, and challenges regarding password management, the goal of this study was to compile suitable advice that can benefit this demographic. To achieve this, we first conducted semi structured interviews (n=37), we then collated advice based on best practice and what we learned from these interviews. We facilitated two independent focus groups (n=31) to evaluate and adjust this advice and tested the finalized advice through an observational study (n=15). The participants were aged between 59 and 86 and came from various counties in Ireland, both rural and urban. 
The findings revealed that managing multiple passwords was a significant source of frustration, leading some participants to adopt novel and informal strategies for storing them. 
A notable hesitation to adopt digital password managers and passphrases was also observed. Participants appreciated guidance on improving their password practices, with many affirming that securely writing down passwords was a practical strategy. Irish older adults demonstrated strong intuition regarding cybersecurity, notably expressing concerns over knowledge-based security checks used by banks and government institutions. 
This study aims to contribute to the aggregation of practical password advice suited to older adults, making password security more manageable and less burdensome for this demographic.

\end{abstract}


%
\IEEEpeerreviewmaketitle

\section{Introduction}
Despite the growing adoption of alternative authentication methods such as biometrics and passkeys, passwords remain the primary form of authentication for most digital services \cite{george2024dawn,lassak2024aren}. Far from becoming obsolete, the use of passwords continues to rise, reflecting their entrenched role in online security systems. Recent data reveals that the average number of passwords per user across online services has increased significantly, from 100 in 2020 to 168 in 2024 \cite{nordpass_passwords_2024}. However, this surge in password use does not necessarily equate to improved cybersecurity practices. Cyberattacks frequently exploit predictable user behaviors, underscoring the gap between password usage and cyber hygiene~\cite{florencio2014password,nisenoff2023two}.

While experts recommend creating strong, unique passwords and avoiding reuse, these best practices often prove challenging for users. Many struggle to remember and manage complex passwords across multiple accounts, creating frustration and fatigue~\cite{adams1999users,herley2009so,florencio2007large}. As digital services continue to expand, these challenges only grow, exacerbating the difficulties of effective password management. Moreover, inconsistencies in security education make it even harder for users to adopt effective strategies~\cite{murray2017evaluating,redmiles2020comprehensive,reeder2017152,lee2022password}, leaving them uncertain about how best to protect their online accounts.

These issues are particularly acute for older adults~\cite{merdenyan2018generational}. A persistent digital divide disproportionately affects this demographic. In Ireland, 65\% of individuals over the age of 65 experience digital exclusion, and 25\% of those aged 60–74 do not use the internet at all \cite{ageaction2021digital}. Compounding this issue, nearly half (47\%) of adults in Ireland lack basic digital skills, according to the Digital Economy and Society Index (DESI) \cite{european2022desi}. These challenges, previously magnified by the COVID-19 pandemic \cite{mccausland2021impact}, are further exacerbated by the rural-urban divide \cite{cso2019urbanrural,flynn2024keeping} and ongoing connectivity issues \cite{mohan2024high}. Together, these factors deepen social isolation and exclusion among older adults while limiting their ability to engage with and understand cybersecurity practices, including password management.

Cybersecurity advice is not a one-size-fits-all solution. Striking a balance between security benefits and usability costs is essential to ensure users can confidently and effectively follow guidance~\cite{redmiles2020comprehensive,murray2017evaluating,herley2009so}. For older adults, this balance is particularly critical. This paper examines the specific password-related concerns and strategies of older adults in Ireland, aiming to develop a prioritized set of guidelines that address their unique needs. Prioritizing advice is crucial, as users have a finite `compliance budget'—the cognitive and practical capacity to adhere to security measures—beyond which they may disengage from best practices altogether~\cite{inglesant2010true,stanton2016security,beautement2008compliance}.

Our research began with an exploration of how Irish older adults navigate password management, focusing on their security concerns and strategies. Through qualitative interviews, we analyzed their approaches to creating and managing passwords, their understanding of best practices, and their barriers to compliance. Findings from this exploratory phase informed a generative phase that developed simplified, accessible password management guidance tailored specifically to this group. Using an iterative design thinking process, we refined this guidance through focus groups and ``think-aloud'' observational studies. This research not only highlights the cybersecurity challenges faced by older adults but also offers practical, user-centered solutions to improve their digital security practices.

\section{Related Work}
A growing body of literature has delved into the challenges older adults face in password management. To begin with, researchers have identified that password creation, updating, and recall present particular challenges for older adults due to cognitive limitations and usability concerns~\cite{stobert2014password,frik2019privacy,mentis2019upside}, often resulting in risky practices such as creating simple passwords~\cite{hargittai2013new,wash2016understanding,frik2019privacy} and reusing the same predictable passwords~\cite{grimes2010older,abela_consumer_2017, wei2024sok}. To address cognitive limitations and usability issues in password management among older adults, cybersecurity experts have long regarded digital password managers as a sure-fire way to streamline password management. However, password manager adoption rates are lowest among older adults \cite{ray2021older}. Reflecting concerns seen in broader populations~\cite{karole2011comparative}, Ray et al.~\cite{ray2021older} reported push-back from older adults exhibiting strong apprehension and mistrust toward cloud storage and cross-device synchronization. They often prefer direct control and tangible methods for password management, such as recording passwords in a notebook and storing it in a secure location. While the adoption of digital password managers may be encouraged through family support and the usability of built-in features like auto-fill, older adults often have incomplete mental models of how password managers work, which can evoke a sense of insecurity within this demographic.

Despite the above evidence of coping mechanisms for password management among older adults internationally, password strategies specific to Irish older adults remain underexplored. To date, only two studies have provided limited qualitative insights into password management strategies, each reporting on the experiences of one older adult. Redahan~\cite{redahan2013older} interviews eight older adults, only one of which discussed passwords. This participant describes the frustrations of using a complex password system on an airline's website, resulting in a time-consuming experience due to frequent password resets and complex requirements. Flynn~\cite{flynn2023ireland} interviewed 20 older adults, with one participant discussing passwords and recalling an instance where a compromised social media password became a source of vulnerability.
Despite threats to other accounts, the participant assumed banking details were secure, indicating an overreliance on institutional safeguards rather than personal password management.


In terms of password creation, Murray and Malone found that Irish users frequently create weak passwords that are culturally and geographically specific, like ``dublin'', ``ireland'', ``munster'' and ``celtic'' \cite{murray2020convergence,murray2023adaptive}. These cultural references are strong indicators of demographic-based password choices.
Nedvěd \cite{nedvved2021careless} notes that the length of words in a language and the use of diacritics can make passwords more unique for example, Irish words are longer compared to English and the use of the Fada (accent) make Irish language (Gaeilge) passwords more secure and unique.
An additional influence on Irish citizens' password habits may be rooted in Ireland's historical context. Garvey and Miller note that Ireland's journey to independence from British colonial rule, along with the subsequent years of social and technological development, has shaped attitudes toward technology \cite{garvey2021ageing}, which we posit may also extend to cybersecurity. As older adults shift from traditional religious frameworks to more individualized and socially integrated lives, they increasingly rely on digital platforms for community and well-being, bringing new concerns over data privacy and cybercrimes.

Technological caregiving from trusted individuals is often necessary, particularly for older adults with mild cognitive impairments~\cite{singh2007password,piper2016technological}. However, this support often involves a trade-off between convenience and security~\cite{latulipe2022unofficial}. Frik et al.~\cite{frik2019privacy} report that many older adults delegate password management to family members or technical assistants, reducing cognitive load but potentially introducing security risks if these helpers lack robust security practices. Mentis et al.~\cite{mentis2020illusion} found that while simplified password practices or shared access with caregivers can mitigate usability issues, these approaches may also heighten vulnerability to cyber threats. Mentis et al suggest however more inclusive approaches to foster meaningful participation of older adults in cybersecurity decisions. This is also reflected in Murthy et al.'s~\cite{murthy2021individually} study, emphasizing the need for designing security solutions that balance collective safety with individual empowerment, ensuring older adults can actively participate in managing their digital security and privacy. 
Only 19.1\% of older adults in Ireland receive technological caregiving from family, below the EU average of 22.3\%~\cite{eurostat2024adulteducation}, this could be due to Ireland's high migration numbers.
In the 12 months leading up to April 2024, over 69,000 individuals emigrated from Ireland, an increase from 64,000 during the same period in 2023. This marks the highest level of emigration recorded since 2015~\cite{Emmigration}. Assuming that most of the Irish older adults rely on their now-emigrated children for help in cybersecurity, these migration patterns suggest that this demographic increasingly resorts to remote family support while also experiencing a sense of forced independence.

This paper builds on the call for region-specific research~\cite{herbert2023world}, by focusing on the unique password management knowledge, perceptions and behaviors of older adults in Ireland. By examining this demographic within a specific cultural and regional context, we aim to contribute to a more detailed understanding of how password management practices and challenges manifest and how education and outreach can best support the needs and requirements of this population. 

\section{Method}
This research forms part of a three-stage study on cyber safety practices conducted in Ireland from November 2023 to August 2024.
This research was approved by our university's institutional review board MTU-HREC-FER-24-009-A. 
Recruitment was carried out through community outreach initiatives, senior centers, and partnerships with digital literacy organizations across the country. While the study primarily targeted individuals aged 65 and over, (65 being the lower bound age for older adults in Europe~\cite{CSOolderadults,eurostat}) if an individual under 65 but aged 50 or older, and self-identifying as an older adult, volunteered to participate at an older adult community event, they were not excluded. As a result, the participant age range for this study was 59 to 86 years.

Participants were provided with an information sheet describing the study before completing a consent form. They were also given the opportunity to ask the researcher interviewing them questions before beginning, and were instructed that they could stop the interview at any time or decline to answer any question. The transcript was recorded live during the interview by a second interviewer. Participants were given the option to consent to audio recording. Where an audio recording was created, it was used to verify the transcript, allowing missed words and phrases to be included. This was not possible for non-recorded interviews, so summaries of participants' answers were sometimes necessary. We found that low-tech users were less likely to agree to being recorded, and we did not want to exclude them from the study. In these cases, a second researcher took detailed notes. These summaries are never presented as direct quotes, and we believe that the value of including all participants in the study outweighed this concession. 
Lastly, participants were compensated with a €40 One4all\footnote{Irish gift cards, \url{https://www.one4all.ie/}} gift voucher to acknowledge their time and contribution.


\textbf{Phase 1: Interviews with Older Adults}
The study began with semi-structured interviews involving 37 (M=11, F=20, Undisclosed=6) older adults across Ireland. The aim of the interview was to explore barriers and concerns related to password use among this demographic. The interview prompts were:
\begin{itemize}
    \item ``What measures do you use to safeguard yourself online?''
    \item ``What's your experience with using passwords?''
    \item ``What's your experience with using verification codes?''
\end{itemize}
These questions allowed participants to guide the discussion based on their experiences and perspectives. We wanted the questions to be open ended and non-leading as much as possible to capture the full picture of password habits among older adults. The interview data was analyzed using Braun and Clarke's reflexive thematic analysis technique~\cite{braun2006using}. This analysis yielded 38 unique codes relating to password use, from which three primary themes emerged: password management, password creation, and password concerns. These themes are discussed in Section~\ref{sec:results}.

\textbf{Phase 2: Co-Creation of Accessible Password Guidance}
Based on the insights gained from the interviews, we compiled simple password guidance informed by best practice password security research~\cite{nist_2025_draft}. This guidance was co-created with 31 older adults through two focus group workshops which took place in Dublin City (urban, n=11, M=5, F=6) and Tipperary Town (rural, n=20, M=2, F=18). The resulting guidance is detailed in Section~\ref{sec:advice}.

\textbf{Phase 3: Observational `Think-Aloud' Study}
The finalized password guidance was then evaluated in a second user study employing a `think aloud' observational approach. Fifteen older adults participated in 1:1 sessions, where they reviewed the written guidance, sharing their thoughts aloud as they went through it. Subsequently, each participant was presented with a hypothetical `Your password has been compromised' notification and asked how they would respond, with the guidance available for reference. This notification was displayed on the researcher's personal computer during the observational session.  Feedback from these sessions is summarized in Section~\ref{sec:feedback}.
The findings are discussed in Section~\ref{sec:disc}, with conclusions presented in Section~\ref{sec:conc}.

\section{Results}\label{sec:results}
In this section, we present the concerns and strategies mentioned by the 37 participants interviewed. Their insights are discussed under the themes of password management, password creation, and password-related concerns. These findings provide the first insights into the password perceptions of Irish older adults. 

\subsection{Password Management}
In speaking with older adults in Ireland, a range of password management practices were uncovered, revealing both innovative and risky strategies. Each time a participant mentioned a security strategy, researchers classified it as secure or insecure based on best practices~\cite{nist_2025_draft}. These findings revealed that 67\% of the password management strategies mentioned by the older adults were secure. 
However, multiple participants expressed frustration at having to remember multiple passwords. Participant 41 declared \textit{``I'm 73 you can't expect me to have multiple passwords!''} 
A number of coping strategies were mentioned in the interviews. One participant chose to store passwords as phone contacts: \textit{``So I'd create a contact. I put it in like it's a phone number, so they think it is''} (P7). When asked about the potential risks of this method, such as someone accessing their contacts, they confidently questioned, \textit{``Why would they be looking at my contacts?''}

Some participants mentioned emailing themselves passwords, while others stored them in notebooks or diaries. For instance, Participant 61 stated, \textit{``Some passwords are memorized, others are in a password book hidden at home''.} Seven of the 37 participants wrote their passwords in a physical book to remember them. A few participants used even more unconventional methods, like renaming files containing passwords with misleading titles. After being admonished by a computer specialist for having a folder on his desktop containing all his passwords under the title `passwords', P19 renamed the file to `rubbish' to obscure its contents. Similar methods of storing passwords have been noted by Pearman et al. \cite{pearman2019people}, where their participants stored passwords in lists on their phones.  While these strategies might seem insecure they are actually reasonable practices. Most attacks will be from an online attacker trying password spraying attacks rather than a targeted attack searching through an individual's phone or checking a local file on their computer \cite{bit-spray}. The same cannot be said for emailing a password to yourself, as this involves transmitting it over the internet and storing it in the cloud, making it vulnerable to online compromise.

The sentiment of nostalgia for a simpler time with fewer passwords was evident. One participant lamented, \textit{``I had a lovely password, and I used it for everything. Then, suddenly, I got a message saying it was compromised. That was the end of my password''} (P19).
Participants had the advice to change their password regularly deeply ingrained in them.  With most seeing regular password changes as the pinnacle of security: \textit{``There is no security; all you can do is keep changing your passwords''} (P6). 
More participants reported regularly changing their passwords than avoiding password reuse across accounts:
\textit{``I reset my password all the time''} (P56). This might suggest that they are changing passwords due to forgetting them, rather than for security reasons. 
Six of the 37 participants mentioned trying not to reuse passwords between different accounts as a security practice they employed \textit{``You try not to use the same password for different accounts''} (P42). Two participants mentioned having different passwords for different types of accounts. Participant 89 said \textit{``I have different passwords for different categories (e.g., banking, email, websites). Depending on the website, that would determine the type of password''} and Participant 69 uses the \textit{``same password for all unimportant stuff and a different password for financial stuff''}. 
Ten participants said they reuse the same password for almost everything. \textit{``I use the same password all the time''} (P63). Participants did realize this was something they shouldn't do, but the effort of creating unique passwords for different accounts was too great: \textit{``I think my biggest thing is using the same password, which I know you shouldn't, but it's very hard to remember so many passwords''} (P96).

\subsection{Password Creation}
When it came to creating passwords, participants exhibited both inventive and problematic approaches. For example, many used personal information, such as dates of birth or wedding anniversaries, as passwords. One participant candidly admitted, \textit{``I use a lot of my DOB, and I know it's very easy to find out''} (P24), while another felt secure using their wedding date: \textit{``I think I still use the day I got married, so they probably wouldn't find that out''} (P25).
Some participants devised more complex methods for password creation, such as combining familiar words with symbols or numbers or incorporating Irish words. Participant 10 described using names of childhood holiday locations, altering them slightly by substituting letters with symbols, to create memorable passwords: \textit{``I use half of one and half of the other, and I might stick in a number and maybe instead of an `s' I put in a dollar sign''} (P10). A few participants opted for automated solutions, such as allowing Google or their iPhone to generate passwords. This introduced different challenges, including not being able to log in the next time they tried and difficulty sharing accounts with their partner.
Advice from family members also influenced password creation strategies, with suggestions ranging from combining random objects in the room to constructing passwords based on poetry. 
These approaches were seen as effective but were not universally adopted.

As verification codes sent by email, SMS or via an app are often used as an additional or alternative authentication mechanism, we also asked participants how they found using these. Surprisingly, participants expressed much less frustration and difficulty with verification codes than they did with passwords. Some participants did however note that the fast expiration times proved to be stressful with one person referring to verification codes as \textit{``A necessary evil''} (P17) and another noting; \textit{``I wouldn't know what to do''} (P57).  Most however felt comfortable with them \textit{``Ok, I have the gist of that''} (P13). In fact, four participants noted that 2FA made them feel safer. 

\subsection{Password concerns}
A recurring theme was the frustration with frequent password changes and the increasing number of passwords to manage. Many participants expressed their dissatisfaction with the process of creating and maintaining passwords, particularly when systems required frequent updates. One participant shared, \textit{``When you set up a password, they allow you to do it, and you spend 10 minutes on it. Then, at the very end, you can't use the password you've just created. That's very frustrating''} (P1).

Participants also voiced concerns about password safety. Several older adults were wary of password managers, either built into their devices or provided by third-party services. Participant 3 expressed mistrust in these systems: \textit{``When you save it to your computer or the cloud, how safe is that? To my mind, it's not safe because I don't have control of it.''} The reluctance to adopt password managers reflects a broader apprehension around trusting technology, compounded by concerns about privacy and control.

Other participants felt that they were of little interest to potential hackers, with one noting, \textit{``I don't have much money in the bank, so if they take it, it's not the end of the world''} (P63). This perception of being a low-value target led some to reuse passwords or create simpler passwords, further underscoring the tension between convenience and security. 

Concerns were raised about the use of standard knowledge-based questions as security checks by banks and government websites, particularly from a genealogy standpoint. Tracing ancestral lineages through historical indexes is a hobby for many older adults. They believe that such official records can provide a gateway for cybercriminals to bypass commonly used security questions. As one participant noted, \textit{``banks ask ridiculous personal questions, in ancestry research a lot of information is available and out there''} (P62). Participant 78 shared a similar sentiment: \textit{``I do genealogy, no problem finding their mother's names or their grandmother's names or their great grandmother's names through genealogy, it's very easy. So the last thing you should do is use your mother's maiden name.''}

\section{Adapted Password Advice}\label{sec:advice}

Drawing from the challenges and practices observed in the interviews, we gathered an initial set of password guidelines aimed specifically at addressing the concerns and needs highlighted by the older adults in this study. We chose to use the NIST Memorized Secret Digital Identity Guidelines as a starting point because they represent international best practices, developed through public consultation and designed with consideration for both usability and security~\cite{nist_2025_draft}.
We presented our initial guidance to older adults in two workshops (n=31; rural=20, urban=11; M=7, F=24). Working collaboratively as a group of experts and non-experts, we co-created useful guidance designed to balance usability with security. 

Password management advice can vary significantly, so it was crucial to strike a balance between providing secure password recommendations and ensuring that the advice remained practical and user-friendly. To achieve this, we iteratively incorporated feedback from the older adults who participated in the workshops. For example, the original guidance did not include advice on sharing passwords, but feedback from the first focus group highlighted this gap: \textit{``You didn't say `never tell another person your password.' It's probably obvious, but, you know, sometimes a person can [ask], you know, what's your [PIN] number there?''}(P59). As a result, we added the advice: `Do not give passwords to anyone over the phone or in person'. However, participants in the second focus group found this advice too restrictive. 
Many older adults mentioned sharing their passwords with a trusted family member to receive help with online tasks.
Compared to younger adults, who often share passwords for convenience, older adults tend to share passwords out of necessity, exercising greater caution and prioritizing trust in the process~\cite{yuan2024account}. This practice is especially common in close-knit or indigenous communities and among people with disabilities~\cite{singh2007password}.  Based on this feedback, the advice was refined to caution against sharing passwords with strangers while acknowledging the possibility of sharing with trusted individuals when necessary. Sharing passwords is often strongly discouraged by cybersecurity experts~\cite{NIST-SHARING-PW}, and many companies explicitly state in their policies that it violates their terms of service or contractual agreements~\cite{FB-PW-SHARING}, potentially breaching regulations such as GDPR or banking standards~\cite{BOI-TOS, GDPR-PW}. However, we suggest that, unless systems are adapted to enable the secure involvement of trusted individuals in account management, these policies should be reconsidered to better address the needs of older adults, who may depend on others to assist with managing their accounts, particularly in cases involving accessibility or cognitive challenges.

Cognitive decline was a key consideration, as it can affect both trust and memory related to technology, making older adults more vulnerable \cite{shang2022psychology,gitlow2014technology,bhattacharjee2020older,butt2023barriers}. Participants expressed concerns about this issue both for themselves and for others they knew. Writing down passwords was suggested as a helpful strategy for those with memory challenges or distrust of digital password managers. As a result, we included guidance on the safe storage of written passwords. For those with greater technological ability, we recommended enabling two-factor authentication (2FA) and using digital password managers as additional options.
Finally, although it was not directly discussed during the initial interviews, we recognized the importance of including advice on what to do if a password is compromised, as this was a significant concern raised during the focus groups.
Below, we outline the finalized advice that was co-created with our workshop participants.


\noindent 
\\
\textbf{Password Management Advice}

\begin{itemize}
    \item Think about your most important accounts. This could be: accounts with your credit card linked to it, your email account, your tax, pension or healthcare accounts. 
    \item For each of these most important accounts create a unique password for each account.
    \item Write these passwords down in a notebook or consider using a digital password manager.
    \item If you trust a family member to keep a password safe for you do so, however this person must bear the responsibility alone and not disclose this password to anyone else. 
    \item Do not give your password to any stranger over the phone or in person.
    \item There is no need to regularly change this password unless you know it has been compromised.
\end{itemize}

\noindent \textbf{Password Creation Advice }
\begin{itemize}
    \item Create a password using a sentence. 
    \begin{itemize}
    \item Recall that spaces count as special characters and can be included to make it easier to remember. 
    \item A sentence should be one you can say easily and if you can it should naturally include special characters and numbers. 
    \item For example: `I make tea at 9:30 am every day'. This password is 31 characters long, includes uppercase and lowercase characters, 3 special characters and 3 numbers.
\end{itemize}
    \item Consider enabling two-factor authentication. This means you will use 2 forms of authentication to access your account (usually a password and a SMS).
\end{itemize}

\noindent\textbf{Password Compromise Advice}
\begin{itemize}
    \item Stay Alert for Notifications
 \begin{itemize}
\item If you receive a notification (email or SMS) warning you about suspicious activity or a compromised password, do not click any links in the message. Instead, go directly to the website and login as normal.
\end{itemize}
    \item Try Logging into Your Account
\begin{itemize}
\item If you can log in, go to your account's Settings and select Change Password.
\item If there's an option to log out of all devices, enable it to secure your account across all platforms.
\item Double-check that the saved recovery email or phone number is your own.
\end{itemize}
    \item If You Are Unable To Log In
\begin{itemize}
 \item Select the `Forgot My Password' option on the login screen to reset your password.
 \item If you don't receive the reset email or message, contact the service provider directly for help regaining access to your account.
 \end{itemize}
    \item Report the Compromise if Necessary
 \begin{itemize}
 \item Reporting the compromise to your local police station may help if you need to provide proof of a breach to the service provider.
 \end{itemize}
\end{itemize}


The above advice was collaboratively developed and refined with older adults as part of our focus group discussions and some further improvements were introduced as part of the think-aloud study detailed below.

\section{Feedback on Advice}
 ~\label{sec:feedback}
Once, the advice was created, we then deployed a one-to-one think-aloud observational study with 15 older adult participants to test the advice.
In the first part of the think-aloud study, participants were given the written advice to read through, expressing their thoughts and reactions as they progressed. They were asked to give a Likert scale 1-5 score for how useful the advice was, receiving an average score of 4.75/5. They were then shown a `Your password has been compromised' notification (Figure~\ref{fig:enter-label}) on the computer. They were asked to follow our advice to reset and create a new password. 
In this section, we will discuss the feedback we received from the think-aloud observational studies. 

\subsection{Password Management Advice Feedback}
\subsubsection{Write it down} Our recommendation to write down passwords in a secure location, such as a notebook stored in a safe place, was well-received: \textit{``Keeping the password written in a secure location is something I'd definitely support''} (P80). Participants seemed relieved that their current practice, often regarded as insecure, was validated. For example, Participant 84 remarked, \textit{``I was told before never to write down your passwords, but I just don't know, unless you write them down, you definitely forget them.''}

\subsubsection{No need to do regular password changes}
Before the observational study, we had not specifically stated that you do not need to change your password regularly. However, when participants were attempting to follow our advice, we realized that we needed to specifically address this misconception as otherwise, participants were trying to combine it with the advice we were giving them: \textit{``There's a lot of management in that. You've got them written down, and you're changing them regularly, right?''} (P95).

\subsubsection{Digital password managers}
Participants were not inclined to use a digital password manager. When asked about it, they instead diverted to talking about their own password storage techniques (in contact lists, sending emails, in files, etc.). Only two participants explicitly said they would not trust a digital password manager: \textit{``If you put your passwords online, I think it's a very bad idea''} (P79), \textit{``Keeping the password written in a secure location is something I'd definitely support, I don't trust electronics''} (P80).
Some participants requested more information on the advice, \textit{``What does that (consider using a digital password manager) mean?''}(P98).

\subsection{Password Creation Advice Feedback}
\subsubsection{Use a sentence as a password}
For password creation, we advised using passphrases—long, memorable phrases that are both secure and easier to recall than arbitrary combinations of characters \cite{shay2014can}. One participant found this approach surprisingly clear: \textit{``I thought you would be avoiding a sentence like that. I was thinking of a name of a place and adding digits''} (P4). However, we also received some concern and surprise about this advice: \textit{``Many websites do not ask for a 3-word sentence''} (P105), 
\textit{``And you have spaces? I didn't realize that actually''} (P98). When tasked with creating a new password one participant (P100) in the observation study declined our advice to use a memorable passphrase on the basis that it would not be secure.
Some participants said that it is one thing to recommend using a password like a sentence, but the restrictions on websites often don't allow your created passwords. \textit{``I'm just surprised about what you said about using the whole sentence; in most cases, you're only allowed to use about 8 digits''} (P107). \textit{``Nearly all the sites say that if you put up something, say your dog's name or something as your password, then come back, and nearly all the sites will come in and say, no, you need a mix of numbers and letters. And they come back and say, this is not strong enough, you need an asterisk, you need whatever. And, of course, you go to try to make up these stupid things. Then you can't remember them, you know''} (P78).
Participants also discussed using the Irish language in their passwords. This can add some additional strength to the passwords as the attacker would need to include an Irish language dictionary of words in their guesses. However, participants were suggesting this as an alternative to a sentence, so we still recommended using a sentence in Irish rather than individual words as these can often still be guessed \cite{murray2023adaptive,maoneke2018influence}.

\subsubsection{Two-factor authentication}
Despite positive feedback on verification codes in the first part of the study, participants were generally unsupportive of two-factor authentication (2FA). This is also reflected in studies looking at 2FA adoption \cite{das2020mfa, golla2021driving}.  In our initial interviews, we used the term `verification codes', while in our guidance, we referred to `two-factor authentication', which may have contributed to this disconnect. Verification codes were perhaps perceived as a routine occurrence, whereas setting up 2FA required active involvement, compounded by the lack of detailed instructions in our advice. For instance, Participant 85 found the 2FA guidance confusing, noting it did not mention the code typically sent during the process.
Additionally, two participants expressed distrust towards 2FA. Participant 90 described it as `dodgy' and was uncomfortable with codes being sent to his phone. Participant 78 noted, \textit{``You tend to be a bit scared anyway, the pressure of 60 seconds… by the time you're thinking should I be doing this in 60 seconds, then it's gone''.} Another participant echoed concerns over time-sensitive codes, saying, \textit{``It can be a bit stressful with 2FA… it's time sensitive, so it pops up, and you have to act within 30 seconds, which creates panic''} (P79).
Several participants also indicated they would appreciate more detailed information on two-factor authentication in our guidance. In future advice we would specify what the different types of 2FA are and how to set them up. 

\subsection{Password Compromise Advice Feedback}

\begin{figure}[h]
    \centering
    \includegraphics[width=0.7\linewidth]{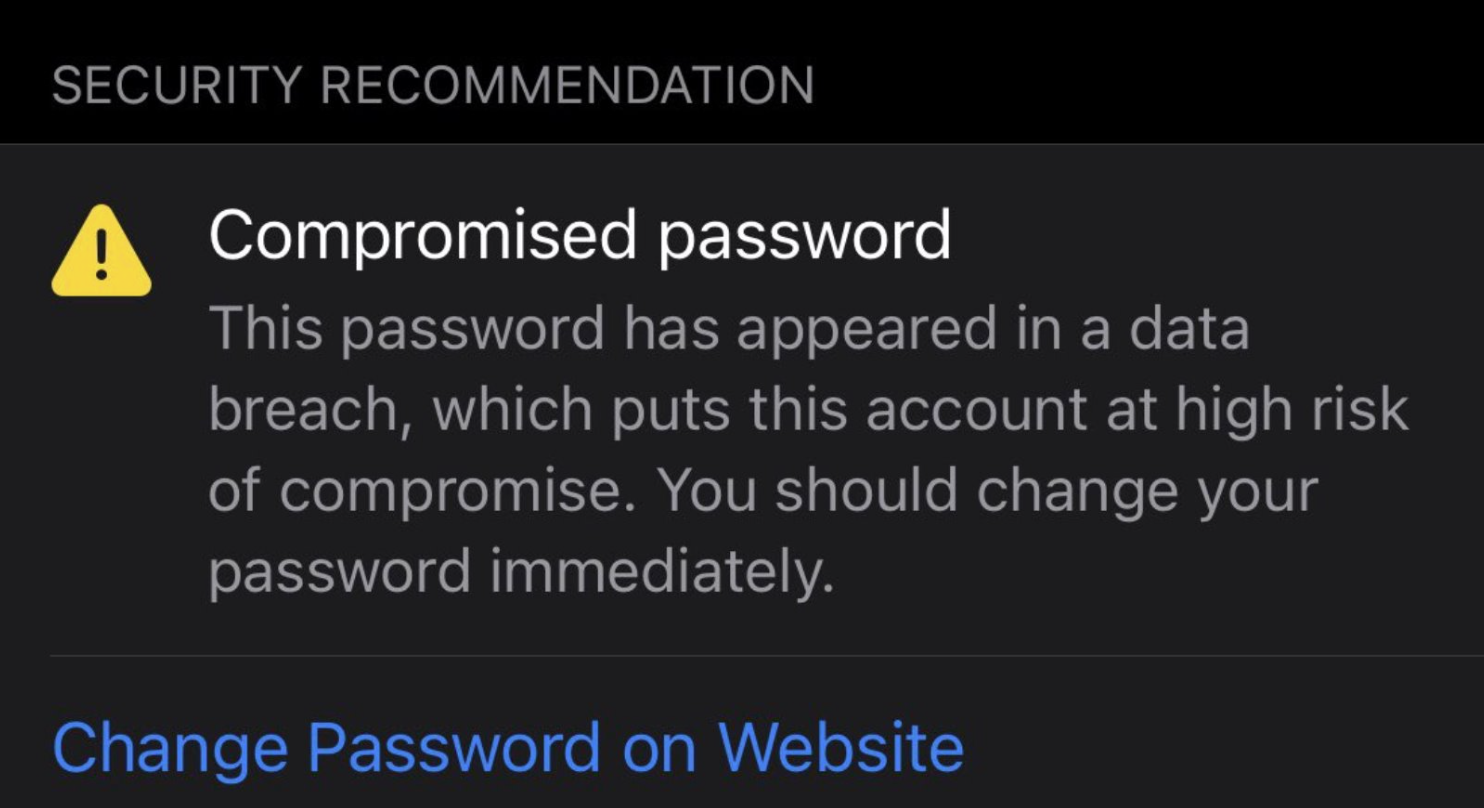}
    \caption{Image prompt for changing password. }
    \label{fig:enter-label}
\end{figure}

In the final part of the study, we showed participants a notification styled like the typical Google Chrome alert, reading, `Your password has been compromised' (Fig.~\ref{fig:enter-label}). We asked them to consider our advice and show us what they would do in this situation.

Three participants said they would ignore the password compromise notification and do nothing about it. \textit{``I would more than likely ignore the message about compromised passwords''} (P87). These participants believed that a message like this would itself be the scam. This echoes the findings of Golla et al. who found less than a third of their respondents reported intentions to change their passwords after seeing a password compromise notification~\cite{golla2018site}. We explained that these notifications can be legitimate but included the advice to not click on it and instead navigate to the website so that they feel more at ease.

Older adults struggled to know how to change the compromised password. Participant 85 was unsure of how to go about changing their password when not using `forgot password'. Older adults also mentioned that advice should address specific steps for changing passwords across different platforms, each with its own recovery and reset processes.

\section{Discussion}
\label{sec:disc}
The findings from this study reveal a significant gap between official password recommendations and the everyday practices of older adults. Although recent cybersecurity guidance emphasizes usability in password advice—focusing on length and memorability over complexity (NIST, 2025; NIST, 2017)—there remains a need to convey this message effectively to the public, especially older adults. 
Despite reporting intuitively secure password practices,
cognitive dissonance emerged as a core theme in the study. Sound password management advice often clashed with previously ingrained beliefs among participants. For example, older adults had long been advised never to write passwords down or to change them regularly. This legacy advice still circulates widely \cite{murray2023costs,redmiles2020comprehensive}, perpetuating insecure practices. The password misconceptions identified in this study are also consistent with those reported in other western countries \cite{herbert2023world}.
Many older adults express overwhelm and frustration regarding password management. Interestingly, some coping strategies they've adopted—such as grouping accounts by importance and using stronger passwords for key accounts—align with recommendations by security researchers like Florêncio et al. \cite{florencio2014password,zhang2016revisiting}. There is a clear need for user-friendly password management solutions that cater to the specific needs of older adults, particularly given potential cognitive decline and lower comfort levels with technology like password managers.

\subsubsection{Password Creation and Storage}
Participants demonstrated both strong (e.g., Irish words with symbols and numbers) and weak (e.g., family members' names) password creation strategies. Notably, storing passwords on a desktop file or in a notebook, while seemingly insecure, is reasonable for older adults who struggle with memory or distrust online password managers~\cite{zhang2016revisiting,boothroyd2013writing}. Given that most cyberattacks are remote \cite{ekele2023keylogger}, targeting exposed datasets rather than individual notebooks, and that personal security habits are generally less risky than workplace security habits \cite{bitwarden_world_password_day}, these practices may suit older adults who find digital alternatives challenging or unaffordable.

\subsubsection{Persistent Challenges and Misconceptions}
\paragraph{Frequent Password Changes} Many participants have a habit of frequently changing passwords, a practice reinforced by usability and memory issues. We advised against this as it increases the risk of password reuse, which can lead to vulnerabilities~\cite{nisenoff2023two}. Tailoring advice to suggest changes only after a compromise or persistent recall issues could make guidance more practical for older adults.

\paragraph{Complex Passwords}
 Participants often default to creating complex, hard-to-remember passwords due to platform requirements. This design choice reflects a broader issue in usability: digital platforms often lack flexibility for older adults, leading to frustrating experiences and discouraging secure practices \cite{shay2016designing,komanduri2011passwords}.

\paragraph{Password Reuse}
 Despite recognizing the importance of unique passwords, many participants continued to reuse them, as the convenience outweighed the perceived security risks. This was also seen in Wei et al.'s Sok paper on the subject of older adults and computer security behaviors \cite{wei2024sok}. The mental relief of using the same password outweighs the abstract risk of hacking attempts across this demographic.

\paragraph{Distrust of Digital Password Managers and Passphrases}

Some participants avoided discussing digital password managers, feeling that their passwords would be more accessible to hackers if stored in browsers. Possible reasons for this distrust is the intangible aspect of digital managers. As older adults often noted the unknown aspects of the digital world, the fact that they cannot see where the passwords are stored plays into this distrust. This mirrors existing literature on distrust in digital technology and password managers among older adults \cite{knowles2018older,ray2021older,seiler2019don}. 
Yet the older adults in our study, appear to be comfortable using the `remember me' function on Google, indicating that convenience plays a role in digital password managers, trust in search engine itself also seems to be a factor. 

Surprisingly, there was also a reluctance to adopt passphrases, as participants associated them with weak strategies, perhaps due to long-standing habits of creating complex strings. This was also seen to be the case in Ur et al.'s user study on password perceptions \cite{ur2016users}. Although Shay et al. \cite{shay2014can} note that with clear guidance passphrases can be more easily adopted.

\subsubsection{Verification Codes and Multi-Factor Authentication}
Verification codes were generally viewed more favorably than passwords, 
although codes with fast expiration times proved stressful for participants.
Older adults expressed frustration at having to act quickly, often requiring multiple attempts. This finding suggests that more accessible authentication methods, such as longer-lasting codes or other alternative verification methods, may be beneficial for older adults. Verification codes seem to be more convenient than setting up multi-factor authentication (MFA), suggesting—similar to the `remember me' feature—that convenience often takes priority over security when it comes to manually configuring security tools like password managers and MFA. 

\subsubsection{Practical Guidance and Foundational Skills}
Although participants found our language and recommendations helpful, many struggled to implement them. For instance, while older adults generally agreed with the need to change passwords when prompted, they required more practical, step-by-step instructions. Providing clear and actionable guidance on changing passwords—both proactively and when prompted—would address a gap in foundational digital skills among older adults.

\subsubsection{Knowledge-Based Authentication}
Older adults recognize the insecurity of using security questions for authentication, yet organizations continue to rely on this outdated practice, particularly in financial and government contexts. Knowledge-based authentication is increasingly viewed as insecure due to its susceptibility to data breaches and the easy accessibility of personal information \cite{grassi_digital_2017, golla2016analyzing}. Older adults' awareness of these risks highlights their intuitive grasp of cybersecurity practices, even as organizations remain slow to adapt.

\subsubsection{Misunderstandings Around Attacks}
Participants often misunderstood how cyberattacks occur. Many assumed they were unlikely to be attacked as there was no reason for them to be targeted. They did not realize that common dictionary-based attacks scan broad databases for easy-to-guess passwords. An introductory explanation of typical password compromise methods could help address this dissonance and improve adherence to secure practices.

\section{Recommendations}
\begin{itemize}
    \item Enhanced Public Awareness Campaigns: Increase the number of public awareness campaigns that offer clear, practical advice on secure practices, grounded in accessible and research-informed security guidance.
    \item Modernize Password Requirements for Users: Organizations need to overhaul outdated password complexity and reset policies, which restrict users and hinder the adoption of more practical, user-centered security practices.
    \item The provision of educational resources about digital password managers to increase comfort and adoption. Additionally focus should be on ways to make them more intuitive and accessible for a broader audience. This same point can be applied to MFA.  
    \item Password sharing can be a necessary requirement for some older adults to complete essential life tasks. However, contracts and recommendations often impose a blanket ban on any form of password sharing. Service providers should offer secure methods for sharing account access, enabling individuals to receive assistance with essential tasks.
\end{itemize}

\section{Limitations}
The study relies primarily on self-reported data, which may not fully represent participants' actual behaviors. Factors such as memory biases, misunderstandings of certain concepts, or a tendency to present themselves in a positive light could impact the data's accuracy.
The password advice presented in this paper is grounded in the password guidelines, managers, and 2FA technologies available in 2024. However, it may not remain applicable as technology continues to evolve.

\section{Conclusion}
 \label{sec:conc}
 This study sheds light on the challenges and adaptations Irish older adults experience in managing passwords. Despite common misconceptions that older adults engage in insecure password practices, we found that the majority are security-conscious, often practicing strategies that align with recommended guidelines. Many older adults found reassurance in practical advice, such as writing down passwords, which validates their intuitive practices and promotes confidence in their ability to manage digital security effectively.
 
Our findings underscore the importance of providing tailored, accessible password guidance that acknowledges the specific needs and habits of older adults. While advice on passphrases and secure storage methods can improve usability, true progress in enhancing password security for this demographic ultimately depends on organizational changes. Organizations must reconsider outdated complexity and reset requirements, as these practices constrain users and limit the effectiveness of more user-friendly security recommendations.
Overall, this research highlights the necessity for collaborative approaches to password management that balance security with usability, creating conditions where all users can manage passwords securely and with less frustration.

\section*{Acknowledgment}
We thank Maximilian Golla, Per Thorsheim and the USEC reviewers for their useful feedback. 
This publication has emanated from research conducted with the financial support of the EU Commission Recovery and Resilience Facility under Research Ireland Our Tech Grant Number 22/NCF/OT/11212G. The research was supported in part by a Google Trust and Safety Research Award. Dr Murray is supported by Taighde Éireann – Research Ireland under Grant number 13/RC/2077\_P2 at CONNECT: the Research Ireland Centre for Future Networks.



%

\bibliographystyle{IEEEtran}
\bibliography{lib}

\begin{thebibliography}{10}
\providecommand{\url}[1]{#1}
\csname url@samestyle\endcsname
\providecommand{\newblock}{\relax}
\providecommand{\bibinfo}[2]{#2}
\providecommand{\BIBentrySTDinterwordspacing}{\spaceskip=0pt\relax}
\providecommand{\BIBentryALTinterwordstretchfactor}{4}
\providecommand{\BIBentryALTinterwordspacing}{\spaceskip=\fontdimen2\font plus
\BIBentryALTinterwordstretchfactor\fontdimen3\font minus \fontdimen4\font\relax}
\providecommand{\BIBforeignlanguage}[2]{{%
\expandafter\ifx\csname l@#1\endcsname\relax
\typeout{** WARNING: IEEEtran.bst: No hyphenation pattern has been}%
\typeout{** loaded for the language `#1'. Using the pattern for}%
\typeout{** the default language instead.}%
\else
\language=\csname l@#1\endcsname
\fi
#2}}
\providecommand{\BIBdecl}{\relax}
\BIBdecl

\bibitem{george2024dawn}
A.~S. George, ``The dawn of passkeys: Evaluating a passwordless future,'' \emph{Partners Universal Innovative Research Publication}, vol.~2, no.~1, pp. 202--220, 2024.

\bibitem{lassak2024aren}
L.~Lassak, E.~Pan, B.~Ur, and M.~Golla, ``Why aren’t we using passkeys? obstacles companies face deploying fido2 passwordless authentication,'' 2024.

\bibitem{nordpass_passwords_2024}
\BIBentryALTinterwordspacing
{NordPass}, ``How many passwords does the average person have?'' 2024, accessed: 2024-11-01. [Online]. Available: \url{https://nordpass.com/blog/how-many-passwords-does-average-person-have/}
\BIBentrySTDinterwordspacing

\bibitem{florencio2014password}
D.~Flor{\^e}ncio, C.~Herley, and P.~C. Van~Oorschot, ``Password portfolios and the $\{$Finite-Effort$\}$ user: Sustainably managing large numbers of accounts,'' in \emph{23rd USENIX Security Symposium (USENIX Security 14)}, 2014, pp. 575--590.

\bibitem{nisenoff2023two}
A.~Nisenoff, M.~Golla, M.~Wei, J.~Hainline, H.~Szymanek, A.~Braun, A.~Hildebrandt, B.~Christensen, D.~Langenberg, and B.~Ur, ``A $\{$Two-Decade$\}$ retrospective analysis of a university's vulnerability to attacks exploiting reused passwords,'' in \emph{{32nd USENIX Security Symposium (USENIX Security 23)}}, 2023, pp. 5127--5144.

\bibitem{adams1999users}
A.~Adams and M.~A. Sasse, ``Users are not the enemy,'' \emph{Communications of the ACM}, vol.~42, no.~12, pp. 40--46, 1999.

\bibitem{herley2009so}
C.~Herley, ``So long, and no thanks for the externalities: the rational rejection of security advice by users,'' in \emph{Proceedings of the 2009 workshop on New security paradigms workshop}, 2009, pp. 133--144.

\bibitem{florencio2007large}
D.~Florencio and C.~Herley, ``A large-scale study of web password habits,'' in \emph{Proceedings of the 16th international conference on World Wide Web}, 2007, pp. 657--666.

\bibitem{murray2017evaluating}
H.~Murray and D.~Malone, ``Evaluating password advice,'' in \emph{2017 28th Irish Signals and Systems Conference (ISSC)}.\hskip 1em plus 0.5em minus 0.4em\relax IEEE, 2017, pp. 1--6.

\bibitem{redmiles2020comprehensive}
E.~M. Redmiles, N.~Warford, A.~Jayanti, A.~Koneru, S.~Kross, M.~Morales, R.~Stevens, and M.~L. Mazurek, ``A comprehensive quality evaluation of security and privacy advice on the web,'' in \emph{29th USENIX Security Symposium (USENIX Security 20)}, 2020, pp. 89--108.

\bibitem{reeder2017152}
R.~W. Reeder, I.~Ion, and S.~Consolvo, ``152 simple steps to stay safe online: Security advice for non-tech-savvy users,'' \emph{IEEE Security \& Privacy}, vol.~15, no.~5, pp. 55--64, 2017.

\bibitem{lee2022password}
K.~Lee, S.~Sj{\"o}berg, and A.~Narayanan, ``Password policies of most top websites fail to follow best practices,'' in \emph{Eighteenth Symposium on Usable Privacy and Security (SOUPS 2022)}, 2022, pp. 561--580.

\bibitem{merdenyan2018generational}
B.~Merdenyan and H.~Petrie, ``Generational differences in password management behaviour,'' in \emph{Proceedings of the 32nd International BCS Human Computer Interaction Conference}.\hskip 1em plus 0.5em minus 0.4em\relax BCS Learning \& Development, 2018.

\bibitem{ageaction2021digital}
A.~Action, ``Digital inclusion and an ageing population,'' 2021, \url{https://www.ageaction.ie/sites/default/files/digital\_inclusion\_and\_an\_ageing\_population.pdf}, Accessed: 7-11-24.

\bibitem{european2022desi}
\BIBentryALTinterwordspacing
E.~Commission, ``The digital economy and society index (desi),'' 2022, accessed: 7-11-14. [Online]. Available: \url{https://digital-strategy.ec.europa.eu/en/policies/desi}
\BIBentrySTDinterwordspacing

\bibitem{mccausland2021impact}
D.~McCausland, R.~Luus, P.~McCallion, E.~Murphy, and M.~McCarron, ``The impact of covid-19 on the social inclusion of older adults with an intellectual disability during the first wave of the pandemic in ireland,'' \emph{Journal of Intellectual Disability Research}, vol.~65, no.~10, pp. 879--889, 2021.

\bibitem{cso2019urbanrural}
\BIBentryALTinterwordspacing
{Central Statistics Office}, ``Urban and rural life in ireland, 2019,'' 2019, available 28 October 2023. [Online]. Available: \url{https://www.cso.ie/en/releasesandpublications/ep/p-urli/urbanandrurallifeinireland2019/introduction/}
\BIBentrySTDinterwordspacing

\bibitem{flynn2024keeping}
S.~Flynn, ``Keeping up with the times in ireland: Older adults bridging the age-based digital divide together?'' \emph{Studies in the Education of Adults}, pp. 1--19, 2024.

\bibitem{mohan2024high}
G.~Mohan and S.~Lyons, ``High-speed broadband availability, internet activity among older people, quality of life and loneliness,'' \emph{new media \& society}, vol.~26, no.~5, pp. 2889--2913, 2024.

\bibitem{inglesant2010true}
P.~G. Inglesant and M.~A. Sasse, ``The true cost of unusable password policies: password use in the wild,'' in \emph{Proceedings of the sigchi conference on human factors in computing systems}, 2010, pp. 383--392.

\bibitem{stanton2016security}
B.~Stanton, M.~F. Theofanos, S.~S. Prettyman, and S.~Furman, ``Security fatigue,'' \emph{It Professional}, vol.~18, no.~5, pp. 26--32, 2016.

\bibitem{beautement2008compliance}
A.~Beautement, M.~A. Sasse, and M.~Wonham, ``The compliance budget: managing security behaviour in organisations,'' in \emph{Proceedings of the 2008 new security paradigms workshop}, 2008, pp. 47--58.

\bibitem{stobert2014password}
E.~Stobert and R.~Biddle, ``A password manager that doesn't remember passwords,'' in \emph{Proceedings of the 2014 New Security Paradigms Workshop}, 2014, pp. 39--52.

\bibitem{frik2019privacy}
A.~Frik, L.~Nurgalieva, J.~Bernd, J.~Lee, F.~Schaub, and S.~Egelman, ``Privacy and security threat models and mitigation strategies of older adults,'' in \emph{Fifteenth symposium on usable privacy and security (SOUPS 2019)}, 2019, pp. 21--40.

\bibitem{mentis2019upside}
H.~M. Mentis, G.~Madjaroff, and A.~K. Massey, ``Upside and downside risk in online security for older adults with mild cognitive impairment,'' in \emph{Proceedings of the 2019 CHI Conference on Human Factors in Computing Systems}, 2019, pp. 1--13.

\bibitem{hargittai2013new}
E.~Hargittai and E.~Litt, ``New strategies for employment? internet skills and online privacy practices during people's job search,'' \emph{IEEE security \& privacy}, vol.~11, no.~3, pp. 38--45, 2013.

\bibitem{wash2016understanding}
R.~Wash, E.~Rader, R.~Berman, and Z.~Wellmer, ``Understanding password choices: How frequently entered passwords are re-used across websites,'' in \emph{Twelfth Symposium on Usable Privacy and Security (SOUPS 2016)}, 2016, pp. 175--188.

\bibitem{grimes2010older}
G.~A. Grimes, M.~G. Hough, E.~Mazur, and M.~L. Signorella, ``Older adults' knowledge of internet hazards,'' \emph{Educational Gerontology}, vol.~36, no.~3, pp. 173--192, 2010.

\bibitem{abela_consumer_2017}
\BIBentryALTinterwordspacing
R.~Abela, ``Consumer survey,'' 2017, accessed: 2024-11-01. [Online]. Available: \url{https://www.netsparker.com/blog/news/consumers-web-applications-most-risk-hacked/}
\BIBentrySTDinterwordspacing

\bibitem{wei2024sok}
M.~Wei, J.~Mink, Y.~Eiger, T.~Kohno, E.~M. Redmiles, and F.~Roesner, ``Sok (or solk?): On the quantitative study of sociodemographic factors and computer security behaviors,'' \emph{arXiv preprint arXiv:2404.10187}, 2024.

\bibitem{ray2021older}
H.~Ray, F.~Wolf, R.~Kuber, and A.~J. Aviv, ``Why older adults (don't) use password managers,'' in \emph{30th USENIX Security Symposium (USENIX Security 21)}, 2021, pp. 73--90.

\bibitem{karole2011comparative}
A.~Karole, N.~Saxena, and N.~Christin, ``A comparative usability evaluation of traditional password managers,'' in \emph{Information Security and Cryptology-ICISC 2010: 13th International Conference, Seoul, Korea, December 1-3, 2010, Revised Selected Papers 13}.\hskip 1em plus 0.5em minus 0.4em\relax Springer, 2011, pp. 233--251.

\bibitem{redahan2013older}
P.~Redahan, ``Older irish adults and the internet. a qualitative study exploring their thoughts, beliefs and practices,'' Ph.D. dissertation, School of Social Work and Social Policy, Trinity College Dublin, 2013.

\bibitem{flynn2023ireland}
S.~Flynn, \emph{Ireland and the Lifelong Learning Curve: The Intergenerational Contribution to Digital Literacy for Life}.\hskip 1em plus 0.5em minus 0.4em\relax Lancaster University (United Kingdom), 2023.

\bibitem{murray2020convergence}
H.~Murray and D.~Malone, ``Convergence of password guessing to optimal success rates,'' \emph{Entropy}, vol.~22, no.~4, p. 378, 2020.

\bibitem{murray2023adaptive}
------, ``Adaptive password guessing: learning language, nationality and dataset source,'' \emph{Annals of Telecommunications}, vol.~78, no.~7, pp. 385--400, 2023.

\bibitem{nedvved2021careless}
V.~Nedv{\v{e}}d, ``Careless society: Drivers of (un) secure passwords,'' 2021.

\bibitem{garvey2021ageing}
P.~Garvey and D.~Miller, \emph{Ageing with smartphones in Ireland: When life becomes craft}.\hskip 1em plus 0.5em minus 0.4em\relax UCL Press, 2021.

\bibitem{singh2007password}
S.~Singh, A.~Cabraal, C.~Demosthenous, G.~Astbrink, and M.~Furlong, ``Password sharing: implications for security design based on social practice,'' in \emph{Proceedings of the SIGCHI conference on Human factors in computing systems}, 2007, pp. 895--904.

\bibitem{piper2016technological}
A.~M. Piper, R.~Cornejo, L.~Hurwitz, and C.~Unumb, ``Technological caregiving: Supporting online activity for adults with cognitive impairments,'' in \emph{Proceedings of the 2016 chi conference on human factors in computing systems}, 2016, pp. 5311--5323.

\bibitem{latulipe2022unofficial}
C.~Latulipe, R.~Dsouza, and M.~Cumbers, ``Unofficial proxies: How close others help older adults with banking,'' in \emph{Proceedings of the 2022 CHI Conference on Human Factors in Computing Systems}, 2022, pp. 1--13.

\bibitem{mentis2020illusion}
H.~M. Mentis, G.~Madjaroff, A.~Massey, and Z.~Trendafilova, ``The illusion of choice in discussing cybersecurity safeguards between older adults with mild cognitive impairment and their caregivers,'' \emph{Proceedings of the ACM on Human-Computer Interaction}, vol.~4, no. CSCW2, pp. 1--19, 2020.

\bibitem{murthy2021individually}
S.~Murthy, K.~S. Bhat, S.~Das, and N.~Kumar, ``Individually vulnerable, collectively safe: The security and privacy practices of households with older adults,'' \emph{Proceedings of the ACM on Human-Computer Interaction}, vol.~5, no. CSCW1, pp. 1--24, 2021.

\bibitem{eurostat2024adulteducation}
Eurostat, ``Adult education survey (aes) - participation in education and training (last 12 months),'' 2024, \url{https://ec.europa.eu/eurostat/databrowser/view/trng\_aes\_201/default/table?lang=en}, Accessed: 7-11-2024.

\bibitem{Emmigration}
{Central Statistics Office}, ``Population and migration estimates, april 2024,'' 2024, \url{https://www.cso.ie/en/releasesandpublications/ep/p-pme/populationandmigrationestimatesapril2024/keyfindings/}, Accessed: 24-11-2024.

\bibitem{herbert2023world}
F.~Herbert, S.~Becker, L.~Schaewitz, J.~Hielscher, M.~Kowalewski, A.~Sasse, Y.~Acar, and M.~D{\"u}rmuth, ``A world full of privacy and security (mis) conceptions? findings of a representative survey in 12 countries,'' in \emph{Proceedings of the 2023 CHI Conference on Human Factors in Computing Systems}, 2023, pp. 1--23.

\bibitem{CSOolderadults}
C.~S. Office, ``Older persons information hub,'' 2024, \url{https://www.cso.ie/en/releasesandpublications/hubs/p-opi/olderpersonsinformationhub}, Acessed: 14-01-25.

\bibitem{eurostat}
Eurostat, ``Ageing europe - introduction,'' 2020, \url{https://ec.europa.eu/eurostat/en/web/products-statistical-books/-/ks-02-20-655}, Acessed: 14-01-25.

\bibitem{braun2006using}
V.~Braun and V.~Clarke, ``Using thematic analysis in psychology,'' \emph{Qualitative research in psychology}, vol.~3, no.~2, pp. 77--101, 2006.

\bibitem{nist_2025_draft}
\BIBentryALTinterwordspacing
D.~Temoshok, D.~Proud-Madruga, Y.-Y. Choong, R.~Galluzzo, S.~Gupta, C.~LaSalle, N.~Lefkovitz, and A.~Regenscheid, ``Digital identity guidelines sp 800-63-4 (2nd public draft),'' National Institute of Standards and Technology (NIST), U.S. Department of Commerce, Tech. Rep., 2024. [Online]. Available: \url{https://csrc.nist.gov/pubs/sp/800/63/4/2pd}
\BIBentrySTDinterwordspacing

\bibitem{pearman2019people}
S.~Pearman, S.~A. Zhang, L.~Bauer, N.~Christin, and L.~F. Cranor, ``Why people (don't) use password managers effectively,'' in \emph{Fifteenth Symposium on Usable Privacy and Security (SOUPS 2019)}, 2019, pp. 319--338.

\bibitem{bit-spray}
Bitwarden, ``How to protect against password spraying attacks,'' \url{https://bitwarden.com/blog/how-to-protect-against-password-spraying-attacks/}, Acessed: 22-11-24.

\bibitem{yuan2024account}
L.~Yuan, Y.~Chen, J.~Tang, and L.~F. Cranor, ``Account password sharing in ordinary situations and emergencies: A comparison between young and older adults,'' in \emph{USENIX Symposium on Usable Privacy and Security (SOUPS), Philadelphia, PA, United States}, 2024.

\bibitem{NIST-SHARING-PW}
\BIBentryALTinterwordspacing
NIST, ``Strength of passwords,'' accessed: 28-01-2025. [Online]. Available: \url{https://pages.nist.gov/800-63-4/sp800-63b.html#appA}
\BIBentrySTDinterwordspacing

\bibitem{FB-PW-SHARING}
\BIBentryALTinterwordspacing
META, ``Meta terms of service,'' accessed: 28-01-2025. [Online]. Available: \url{https://www.facebook.com/legal/terms}
\BIBentrySTDinterwordspacing

\bibitem{BOI-TOS}
\BIBentryALTinterwordspacing
B.~of~Ireland, ``Terms of use of the bank of ireland mortgage solution (``website''),'' accessed: 28-01-2025. [Online]. Available: \url{https://mortgagesolutionsfs.bankofireland.com/Resources/PDF/termsofuse.pdf}
\BIBentrySTDinterwordspacing

\bibitem{GDPR-PW}
\BIBentryALTinterwordspacing
D.~P. Commission, ``Guidance for controllers on data security,'' accessed: 28-01-2025. [Online]. Available: \url{www.dataprotection.ie/sites/default/files/uploads/2019-06/190625%20Data%20Security%20Guidance.pdf}
\BIBentrySTDinterwordspacing

\bibitem{shang2022psychology}
Y.~Shang, Z.~Wu, X.~Du, Y.~Jiang, B.~Ma, and M.~Chi, ``The psychology of the internet fraud victimization of older adults: A systematic review,'' \emph{Frontiers in psychology}, vol.~13, p. 912242, 2022.

\bibitem{gitlow2014technology}
L.~Gitlow, ``Technology use by older adults and barriers to using technology,'' \emph{Physical \& Occupational Therapy in Geriatrics}, vol.~32, no.~3, pp. 271--280, 2014.

\bibitem{bhattacharjee2020older}
P.~Bhattacharjee, S.~Baker, and J.~Waycott, ``Older adults and their acquisition of digital skills: A review of current research evidence,'' in \emph{Proceedings of the 32nd Australian conference on human-computer Interaction}, 2020, pp. 437--443.

\bibitem{butt2023barriers}
S.~A. Butt, S.~Lips, R.~Sharma, I.~Pappel, and D.~Draheim, ``Barriers to digital transformation of the silver economy: Challenges to adopting digital skills by the silver generation,'' \emph{Proceedings of AHFE}, pp. 151--163, 2023.

\bibitem{shay2014can}
R.~Shay, S.~Komanduri, A.~L. Durity, P.~Huh, M.~L. Mazurek, S.~M. Segreti, B.~Ur, L.~Bauer, N.~Christin, and L.~F. Cranor, ``Can long passwords be secure and usable?'' in \emph{Proceedings of the SIGCHI Conference on Human Factors in Computing Systems}, 2014, pp. 2927--2936.

\bibitem{maoneke2018influence}
P.~B. Maoneke, S.~Flowerday, and N.~Isabirye, ``The influence of native language on password composition and security: A socioculture theoretical view,'' in \emph{ICT Systems Security and Privacy Protection: 33rd IFIP TC 11 International Conference, SEC 2018, Held at the 24th IFIP World Computer Congress, WCC 2018, Poznan, Poland, September 18-20, 2018, Proceedings 33}.\hskip 1em plus 0.5em minus 0.4em\relax Springer, 2018, pp. 33--46.

\bibitem{das2020mfa}
S.~Das, B.~Wang, A.~Kim, and L.~J. Camp, ``Mfa is a necessary chore!: Exploring user mental models of multi-factor authentication technologies.'' in \emph{HICSS}, 2020, pp. 1--10.

\bibitem{golla2021driving}
M.~Golla, G.~Ho, M.~Lohmus, M.~Pulluri, and E.~M. Redmiles, ``Driving $\{$2FA$\}$ adoption at scale: Optimizing $\{$Two-Factor$\}$ authentication notification design patterns,'' in \emph{30th USENIX Security Symposium (USENIX Security 21)}, 2021, pp. 109--126.

\bibitem{golla2018site}
M.~Golla, M.~Wei, J.~Hainline, L.~Filipe, M.~D{\"u}rmuth, E.~Redmiles, and B.~Ur, ``" what was that site doing with my facebook password?" designing password-reuse notifications,'' in \emph{Proceedings of the 2018 ACM SIGSAC Conference on Computer and Communications Security}, 2018, pp. 1549--1566.

\bibitem{murray2023costs}
H.~Murray and D.~Malone, ``Costs and benefits of authentication advice,'' \emph{ACM Transactions on Privacy and Security}, vol.~26, no.~3, pp. 1--35, 2023.

\bibitem{zhang2016revisiting}
L.~Zhang-Kennedy, S.~Chiasson, and P.~van Oorschot, ``Revisiting password rules: facilitating human management of passwords,'' in \emph{2016 APWG symposium on electronic crime research (eCrime)}.\hskip 1em plus 0.5em minus 0.4em\relax IEEE, 2016, pp. 1--10.

\bibitem{boothroyd2013writing}
V.~Boothroyd and S.~Chiasson, ``Writing down your password: Does it help?'' in \emph{2013 Eleventh Annual Conference on Privacy, Security and Trust}.\hskip 1em plus 0.5em minus 0.4em\relax IEEE, 2013, pp. 267--274.

\bibitem{ekele2023keylogger}
C.~Ekele~Victoria, A.~Adebiyi~Ayodele, and O.~Igbekele~Emmanuel, ``Keylogger detection: A systematic review,'' in \emph{International Conference on Science, Engineering and Business for Sustainable Development Goals (SEB-SDG)}, 2023, pp. 1--6.

\bibitem{bitwarden_world_password_day}
Bitwarden, ``World password day,'' \url{https://bitwarden.com/resources/world-password-day/}, 2024, accessed: 2024-11-25.

\bibitem{shay2016designing}
R.~Shay, S.~Komanduri, A.~L. Durity, P.~Huh, M.~L. Mazurek, S.~M. Segreti, B.~Ur, L.~Bauer, N.~Christin, and L.~F. Cranor, ``Designing password policies for strength and usability,'' \emph{ACM Transactions on Information and System Security (TISSEC)}, vol.~18, no.~4, pp. 1--34, 2016.

\bibitem{komanduri2011passwords}
S.~Komanduri, R.~Shay, P.~G. Kelley, M.~L. Mazurek, L.~Bauer, N.~Christin, L.~F. Cranor, and S.~Egelman, ``Of passwords and people: measuring the effect of password-composition policies,'' in \emph{Proceedings of the sigchi conference on human factors in computing systems}, 2011, pp. 2595--2604.

\bibitem{knowles2018older}
B.~Knowles and V.~L. Hanson, ``Older adults’ deployment of ‘distrust’,'' \emph{ACM Transactions on Computer-Human Interaction (TOCHI)}, vol.~25, no.~4, pp. 1--25, 2018.

\bibitem{seiler2019don}
S.~Seiler-Hwang, P.~Arias-Cabarcos, A.~Mar{\'\i}n, F.~Almenares, D.~D{\'\i}az-S{\'a}nchez, and C.~Becker, ``" i don't see why i would ever want to use it" analyzing the usability of popular smartphone password managers,'' in \emph{Proceedings of the 2019 ACM SIGSAC Conference on Computer and Communications Security}, 2019, pp. 1937--1953.

\bibitem{ur2016users}
B.~Ur, J.~Bees, S.~M. Segreti, L.~Bauer, N.~Christin, and L.~F. Cranor, ``Do users' perceptions of password security match reality?'' in \emph{Proceedings of the 2016 CHI conference on human factors in computing systems}, 2016, pp. 3748--3760.

\bibitem{grassi_digital_2017}
\BIBentryALTinterwordspacing
P.~A. Grassi, M.~E. Garcia, and J.~L. Fenton, ``Digital identity guidelines sp 800-63-3,'' National Institute of Standards and Technology (NIST), U.S. Department of Commerce, Tech. Rep., 2017, includes updates as of 03-02-2020. [Online]. Available: \url{https://doi.org/10.6028/NIST.SP.800-63-3}
\BIBentrySTDinterwordspacing

\bibitem{golla2016analyzing}
M.~Golla and M.~D{\"u}rmuth, ``Analyzing 4 million real-world personal knowledge questions (short paper),'' in \emph{Technology and Practice of Passwords: 9th International Conference, PASSWORDS 2015, Cambridge, UK, December 7--9, 2015, Proceedings 9}.\hskip 1em plus 0.5em minus 0.4em\relax Springer, 2016, pp. 39--44.

\end{thebibliography}
\end{document}